\documentclass[aps,twocolumn,preprintnumbers,showpacs,showkeys,nofootinbib%
]{revtex4}
\usepackage{epsfig}
\usepackage{amssymb,amsmath,amsfonts,amsthm,graphicx,psfrag}
\graphicspath{{./Figures/}}                 
\usepackage{array}
\usepackage{picture}
\usepackage{subfig}
\usepackage{graphicx}
\usepackage{textpos}
\usepackage{color}

\newcommand{\beq}{\begin{equation}}
\newcommand{\eeq}{\end{equation}}
\newcommand{\bea}{\begin{eqnarray}}
\newcommand{\eea}{\end{eqnarray}}
\newcommand{\aabf}{{\mathbf{A}}}

\newcommand{\re}{\operatorname{\mathfrak{Re}}}
\newcommand{\tr}{\operatorname{T\!r}}

\newcommand{\ds}{\displaystyle}

\newcommand{\aasymt}{{\cal A}}

\begin{document}

\title{The Difference between the Longitudinal and Transverse Gluon Propagators
as an Indicator of the Postconfinement Domain.}

\author{V.~G.~Bornyakov}
\affiliation{Institute for High Energy Physics NRC ``Kurchatov Institute'', 142281 Protvino, Russia \\
Institute of Theoretical and Experimental Physics NRC ``Kurchatov Institute'', 117218 Moscow, Russia}

\author{N.~V.~Gerasimeniuk}
\affiliation{Pacific Quantum Center, Far Eastern Federal University, 690950 Vladivostok, Russia}

\author{V.~A.~Goy}
\affiliation{Pacific Quantum Center, Far Eastern Federal University, 690950 Vladivostok, Russia \\
Institute of Theoretical and Experimental Physics NRC ``Kurchatov Institute'', 117218 Moscow, Russia}

\author{R.~N.~Rogalyov}
\affiliation{Institute for High Energy Physics NRC ``Kurchatov Institute'', 142281 Protvino, Russia \\
}


\begin{abstract}
We study numerically the dependence of the difference
between the longitudinal and transverse gluon propagators,
$\Delta=D_L-D_T$, on the momentum and temperature at $T\gtrsim T_c$ both in SU(2) and SU(3) gluodynamics.
It is found that the integral of $\Delta$ with respect to the 3-momentum is sensitive only to infrared dynamics
and shows a substantial correlation with the Polyakov loop.
At $T=T_p\sim 1.2 T_c$ it changes sign
giving some evidence that $T_p$ can serve as a boundary of the postconfinement domain.
\end{abstract}



\maketitle

\section{Introduction}
\label{sec:introduction}

In the last decades great attention has been given to the theoretical and experimental studies of strong-interacting matter at high temperatures. It was found that at $T\sim 145\div163$~MeV quark-gluon matter undergoes a crossover transition to chirally symmetric deconfinement phase \cite{Bazavov:2011nk,HotQCD:2014kol}. 
However, our understanding of the nature of these  transitions is far from being complete.
Our study focuses on the deconfinement transition. 

One of conventional tools for investigating 
the deconfinement transition is to consider
OCD-like theories, which are simpler than the QCD 
but 
also have such transition.
One of such theories is the heavy-quark limit of QCD, 
in which the fermion degrees of freedom can be ignored
and one arrives at gluodynamics, which is described by
SU(3) pure gauge theory.
In this limit, the crossover transition turns into
the first-order phase transition and the Polyakov loop
provides an order parameter of this transition.
However, the renormalized Polyakov loop
jumps from zero to only $\sim$0.4 at $T=T_c$ and then 
increases with temperature over the range $T_c < T < 4T_c$
\cite{Kaczmarek:2002mc,Dumitru:2003hp}.
In some earlier works, this interval of temperatures 
was referred to as the postconfinement domain
and strongly interacting matter 
in this temperature range
— as semi-QGP \cite{Dumitru:2010mj,Hidaka:2015ima}.
At these temperatures the quark-gluon matter 
demonstrates special properties.  
In particular, the pressure of the semi-QGP differs substantially from the ideal-gas value.
Yet another evidence for the validity of the 
"postconfinement domain" concept
is the prediction \cite{Asakawa:2003re} that heavy quarkonia survive up to
$T = 1.6T_c$. 

Another QCD-like theory worth exploring is the SU(2) pure gauge theory, though the deconfinement phase transition in SU(2) gluodynamics is of the second order.

The objects which are considered to be deconfined
in gluodynamics are the gluons, that is, quanta of the 
gauge field. For this reason, studies of the correlation functions of the gluon fields are of primary importance
on a way to understanding of both the mechanism of confinement and the transition to deconfinement.

We study the behavior of the Landau-gauge gluon propagators  at $T\gtrsim T_c$ in the SU(2) and SU(3) gluodynamics with a particular emphasis on the dependence of the difference between the longitudinal and transverse propagators on 
the momentum and the temperature.
We find an exponential decrease of this difference
in a sufficiently wide range of momenta.
This finding indicates that the dominating contribution 
to the integral 
$$
\Xi = \int d\vec p (D_L(\vec p, 0) - D_T(\vec p, 0))
$$
comes from the infrared domain.
We study the temperature dependence
of $\Xi$ and find that it behaves differently in different Polyakov-loop sectors (center sectors) and,
in the sector with a positive value of the real part of the Polyakov loop 
it changes sign at the temperature $T=T_p\sim 1.2T_c$.
We discuss the relation of this temperature to the boundary of the postconfinement domain.

\section{Definitions and simulation details}

We study SU(2) and 
SU(3) lattice gauge theories with 
the standard Wilson action in the Landau gauge. 
Definitions of the chromo-electric-magnetic 
asymmetry and the propagators
can be found e.g. in 
\cite{Chernodub:2008kf,Bornyakov:2016geh,Bornyakov:2011jm,Aouane:2011fv}.


Link variable $U_{x\mu}$ is related to the Yang-Mills 
vector potential $A_\mu^b(\vec x, x_4)$ as follows.
One determines a Hermitian traceless matrix 
\beq
z={1\over 2\imath}\left( U_{x\mu}-U_{x\mu}^\dagger - {1\over N_c}\,\tr\big( U_{x\mu}-U_{x\mu}^\dagger\big)  \right)
\eeq
which is connected with the dimensionless vector potentials ($a$ is the lattice spacing)
\beq
u^b_\mu(x) = \;-\; {ga\over 2} A_\mu^b(x)\;  ,
\eeq
by the formulas
\beq
z_{ij}= u^b_\mu(x) \Gamma^b_{ij}, \qquad u^b_\mu(x) = 2\,\mathrm{T\!r} \Big(\Gamma^b z\Big) = 2 \Gamma^b_{ij} z_{ji}\,,
\eeq 
where $\Gamma^a$ are Hermitian generators of $SU(N_c)$ 
normalized so that
\beq
\langle \Gamma^a \Gamma^b \rangle \equiv \mathrm{T\!r} (\Gamma^a \Gamma^b) = \Gamma^a_{ij} \Gamma^b_{ji} = {1\over 2} \delta^{ab}
\eeq

In the fundamental representation,
\begin{displaymath}
\Gamma^a = \left[ \begin{array}{lcl}
                     \ds {\sigma^a \over 2}    & \mbox{~~for} & SU(2) \\[3mm]
                     \ds {\lambda^a \over 2}   & \mbox{~~for} & SU(3) 
                    \end{array} \right.
\end{displaymath}

Transformation of the link variables $U_{x\mu}\in SU(3)$ 
under gauge transformations $g_x \in SU(3)$ has the form 
$$ U_{x\mu}
\stackrel{g}{\mapsto} U_{x\mu}^{g} = g_x^{\dagger} U_{x\mu} g_{x+\mu}\;.
$$

The lattice Landau gauge condition is given by
\beq
(\partial \aabf)_{x} = \sum_{\mu=1}^4 \left( \aabf_{x\mu}
- \aabf_{x-\hat{\mu};\mu} \right)  = 0 \,.
\eeq
It represents a stationarity condition for the gauge-fixing functional
\beq\label{eq:gaugefunctional}
F_U(g) = \frac{1}{4V}\sum_{x,\mu}~\frac{1}{3}~\re\tr~U^{g}_{x\mu} \;,
\eeq
with respect to gauge transformations $g_x~$.

\vspace*{2mm}

Our calculations are performed on asymmetric lattices $N_t\times N_s^3$,
where $N_t$ is the number of sites in the temporal direction. In our study, $N_t=8$, $N_s = 24$ in the case of SU(3) and $N_t=8$, $N_s$ varies so that $L=N_sa\approx 3$~fm in the case of SU(2). 
The physical momenta $p$ are given by $\hat p_i=\big(2/a\big) \sin{(\pi
k_i/N_s)}, ~~\hat p_{4}=(2/a) \sin{(\pi k_4/N_t)}, ~~k_i \in (-N_s/2,N_s/2], k_4 \in
(-N_t/2,N_t/2]$. We consider only soft modes $p_4=0$. 

The temperature $T$ is given by $~T=1/aN_t~$ where $a$
is the lattice spacing determined by the coupling constant. We use the parameter
\beq
\tau = {T-T_c \over T_c}
\eeq 
at temperatures close to $T_c$.

In the SU(3) case, we rely on the scale fixing procedure proposed in \cite{Necco:2001xg}
and use the value of the Sommer parameter $r_0=0.5$~fm as in \cite{Bornyakov:2011jm}. 
Making use of $\beta_c=6.06$ and 
$\ds {T_c\over \sqrt{\sigma}}=0.63$
 \cite{Boyd:1996bx} gives $T_c=294$~MeV and 
$\sqrt{\sigma}=0.47$~GeV.

In the SU(2) case we find the relation between lattice spacing $a$ and lattice coupling  $\beta$
from a fit to the lattice data \cite{Fingberg:1992ju} for 
$a\sqrt{\sigma}$ for some set values of $\beta$, where  $\sigma=(440$~MeV$)^2$ is the string tension.

We provide information on lattice spacings, temperatures and other parameters used in this work
in Tables~\ref{tab:statistics_SU2} (SU(2)) 
and~\ref{tab:statistics_SU3} (SU(3)).

\begin{table}[tbh]
\begin{center}
\vspace*{0.2cm}
\begin{tabular}{|c|c|c|c|c|} \hline
           &          &               &                &          \\[-2mm]
 ~~~~$\beta$~~~~ & ~~$a$~fm~ & $a^{-1}$,~GeV & $T$,~MeV & ~~~~$\tau$~~~~ \\[-2mm]
           &          &               &                &          \\
   \hline\hline
2.478 & 0.0921 & 2.143 & 267.9 & -0.099 \\
2.508 & 0.0836 & 2.359 & 294.9 & -0.0077 \\
2.510 & 0.0831 & 2.374 & 296.8 & -0.0013 \\
2.513 & 0.0823 & 2.397 & 299.7 & 0.0083 \\
2.515 & 0.0818 & 2.412 & 301.6 & 0.0148 \\
2.521 & 0.0802 & 2.459 & 307.4 & 0.0345 \\
2.527 & 0.0787 & 2.507 & 313.4 & 0.0545 \\
2.542 & 0.0750 & 2.631 & 328.8 & 0.106 \\
2.547 & 0.0738 & 2.672 & 334.0 & 0.123 \\
2.552 & 0.0727 & 2.715 & 339.4 & 0.141 \\
2.557 & 0.0714 & 2.762 & 345.3 & 0.160 \\
2.562 & 0.0704 & 2.802 & 350.3 & 0.178 \\
2.567 & 0.0693 & 2.847 & 355.9 & 0.198 \\
2.572 & 0.0682 & 2.894 & 361.7 & 0.217 \\
2.586 & 0.0652 & 3.025 & 378.2 & 0.272 \\
2.600 & 0.0624 & 3.157 & 394.6 & 0.329 \\
2.637 & 0.0556 & 3.551 & 443.9 & 0.494 \\
2.701 & 0.0455 & 4.341 & 542.6 & 0.825 \\
2.779 & 0.0357 & 5.524 & 690.6 & 1.325 \\
\hline\hline
\end{tabular}
\end{center}
\caption{Parameters associated with lattices under study
}
\label{tab:statistics_SU2}
\end{table}

\begin{table}[tbh]
\begin{center}
\vspace*{0.2cm}
\begin{tabular}{|c|c|c|c|c|c|} \hline
           &          &               &       &         &          \\[-2mm]
 ~~~~$\beta$~~~~ & ~~$a$~fm~ & $a^{-1}$,~GeV &  $p_{min}$,~MeV & ~~~~$\tau$~~~~ &~$N_\tau$~ \\[-2mm]
           &   &          &               &                &          \\
   \hline\hline
 6.000 & 0.093 & 2.118 & 554.5 & -0.096 & 8\\
 6.044 & 0.086 & 2.283 & 597.7 & -0.026 & 8\\
 6.075 & 0.082 & 2.402 & 628.8 &  0.025 & 8\\
 6.122 & 0.076 & 2.588 & 677.5 &  0.104 & 8\\
 5.994 & 0.098 & 2.096 & 470.1 & 0.192 & 6 \\
\hline\hline
\end{tabular}
\end{center}
\caption{Parameters associated with lattices under study
}
\label{tab:statistics_SU3}
\end{table}

In the SU(2) case we generate $\sim$1000
independent Monte Carlo
gauge-field configurations for each temperature under consideration so that at $\tau< 0.015$ both Polyakov-loop sectors are taken into consideration,
at higher temperatures we study only the sector with ${\cal P}>0$. 

We vary 
lattice sizes at $T=1.015T_c$
and at $T=1.5T_c$ in order to estimate finite-volume effects; in other cases lattice size is $\approx 3$~fm.

In the SU(3) case, 
we generate ensembles of $200$ configurations 
for each of the sectors:
\bea
(I)   \qquad -\;{\pi\over 3} < &\arg {\cal P}& < {\pi\over 3}   \\ \nonumber
(II)  \qquad\quad    {\pi\over 3} < &\arg {\cal P}& < \pi   \\ \nonumber
(III) \qquad -\;{\pi}        < &\arg {\cal P}& < -\;{\pi\over 3} \;   \nonumber
\eea
in order to consider all three Polyakov-loop sectors in detail (${\cal P}$ is the Polyakov loop).
Consecutive configurations (considered as 
independent) were separated by $200\div 400$ 
sweeps, each sweep consisting of one local heatbath update followed by $N_s/2$ microcanonical updates. 

Following Refs.~\cite{Bornyakov:2011jm,Aouane:2011fv} 
we use the gauge-fixing algorithm 
that combines $Z(3)$ flips for space directions with the simulated annealing 
(SA) algorithm followed by overrelaxation. 

Here we do not consider details of the approach to the continuum limit  and renormalization considering that the lattices with $N_t=8$ 
(corresponding to spacing $a\simeq 0.08$~fm at $T \sim T_c$) are sufficiently fine.

We also consider the chromoelectric-chromomagnetic
asymmetry \cite{Chernodub:2008kf,Bornyakov:2016geh}
as an indicator of the relative strength of
chromoelectric and chromomagnetic interactions and  compare it with the Integrated Difference of Propagators
(IDP) $\Xi(T)$ introduced in this work (see eq.~(\ref{eq:IDP_def}) below). In terms of lattice variables, the asymmetry has the form
\beq\label{eq:asym_thr_prop}
\aasymt={6 a^2 N_t^2\over \beta }
\sum_{b=1}^8\left( \Big\langle A_{x,4}^b A_{x,4}^b \Big\rangle - {1\over 3}\sum_{i=1}^3
 \Big\langle A_{x,i}^b A_{x,i}^b \Big\rangle \right),
\eeq
It can also be expressed in terms of the gluon propagators:
\bea\label{eq:average_bare_asymmetry}
&& \aasymt  = {16 N_t \over \beta a^2 N_s^3}
\Big[ 3 (D_L(0) - D_T(0))  \\ \nonumber
&+& \sum_{p\neq 0}\left(
{3|\vec p|^2 \,-\, p_4^2\over p^2}  D_L(p) - 2  D_T(p)\right)\Big] \nonumber
\eea
where $D_L (D_T)$ is the longitudinal (transversal) gluon propagator.
Thus the asymmetry $\aasymt$, which is nothing but
the vacuum expectation value of the respective composite operator,
is multiplicatively renormalizable and its renormalization factor
coincides with that of the propagator\footnote{Assuming
that both $D_L(p)$ and $D_T(p)$ are renormalized by the same factor.}.
\vspace{2mm}

\section{Momentum dependence of $D_L-D_T$ }

\begin{figure*}[bht]
\vspace*{-16mm}
\hspace*{-5mm}\includegraphics[width=9cm]{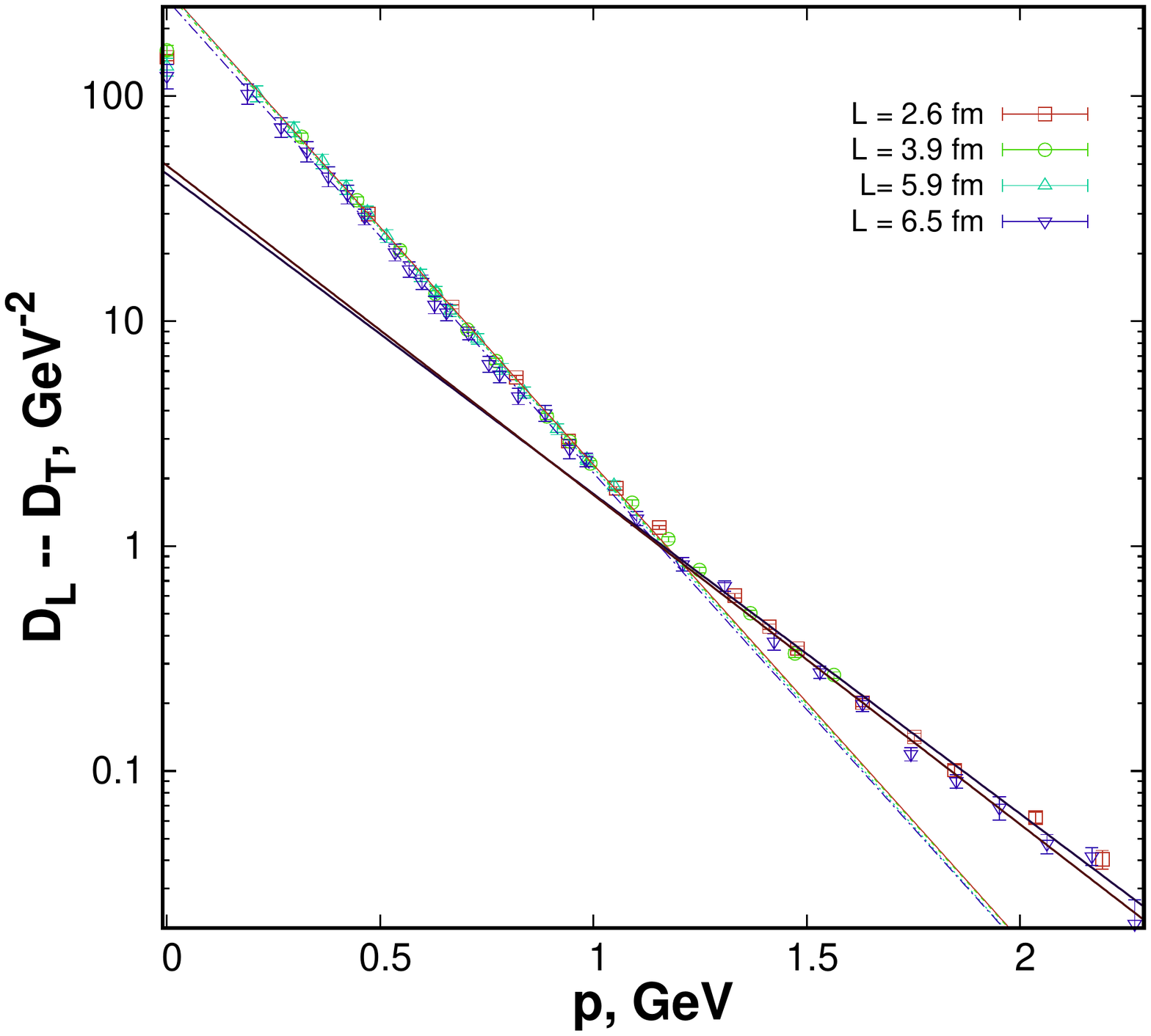}\hspace*{-8mm}
\includegraphics[width=9cm]{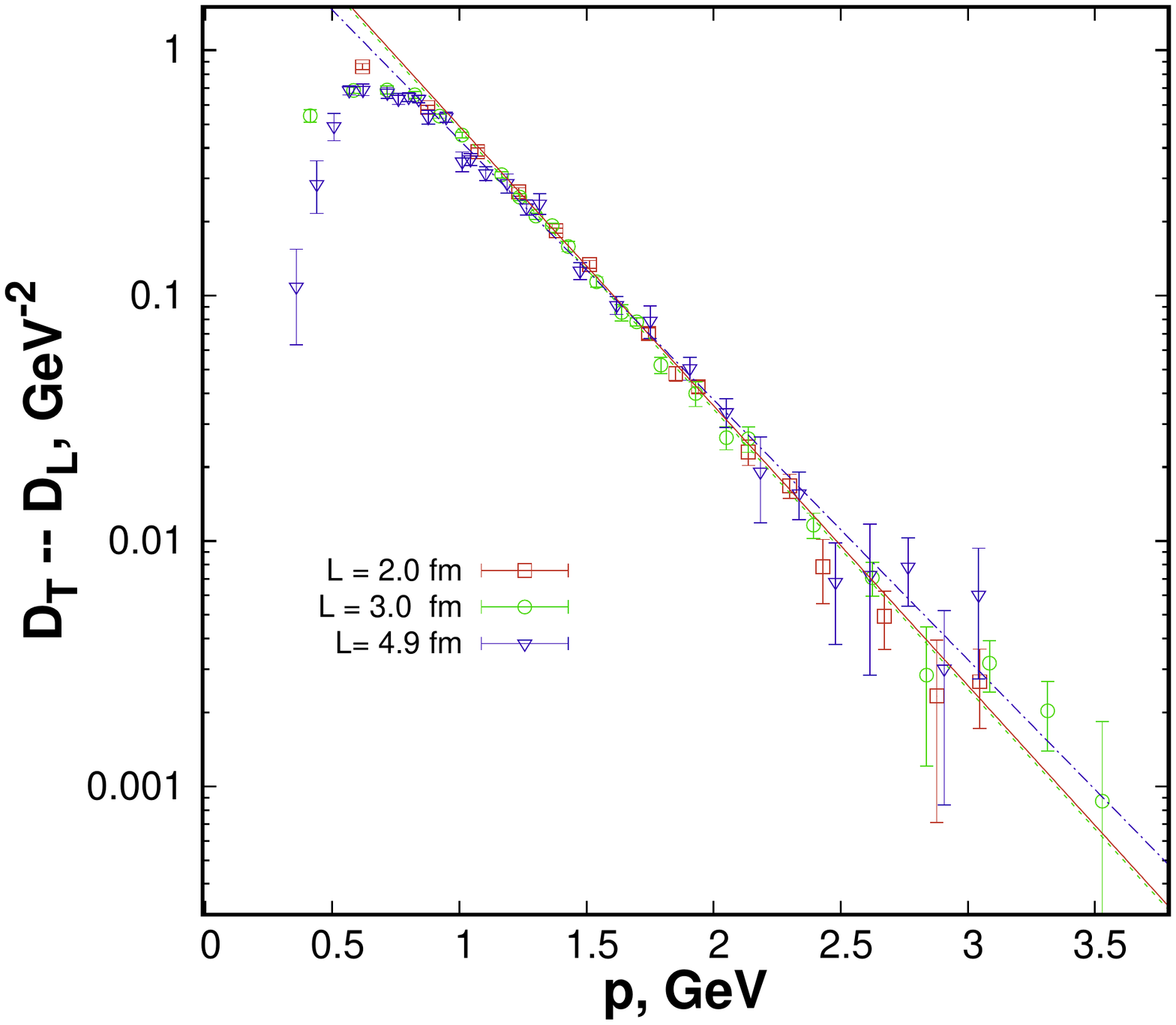}
\vspace*{-21mm}
\caption{Dependence of $\Delta$ on $|\vec p|$ in the SU(2) theory  
at $T=1.014T_c$ (left panel) 
and $T=1.5 T_c$ (right panel). Lines are the results of the fit by eq.~(\ref{eq:diff_basic_fit}).
In the left panel we show the results of fitting 
by eq.~(\ref{eq:diff_basic_fit}) over two ranges: $0.2<|\vec p|<1.2$~GeV and $1.3<|\vec p|<2.2$~GeV.}
\label{fig:FVE}
\end{figure*}
\begin{figure*}[tbh]
\vspace*{1mm}
\hspace*{-5mm}\includegraphics[width=9cm]{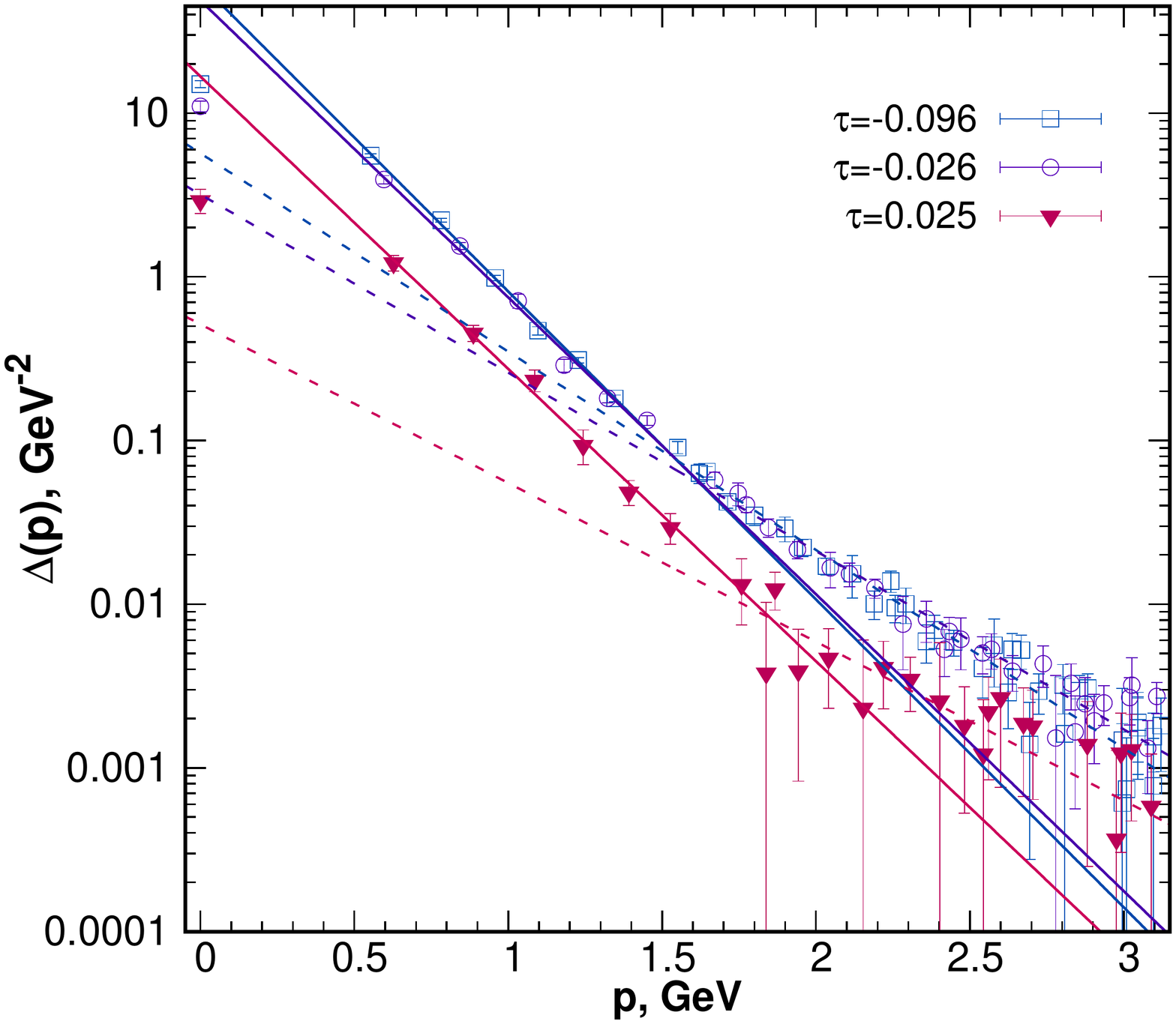}\hspace*{-8mm}
\includegraphics[width=9cm]{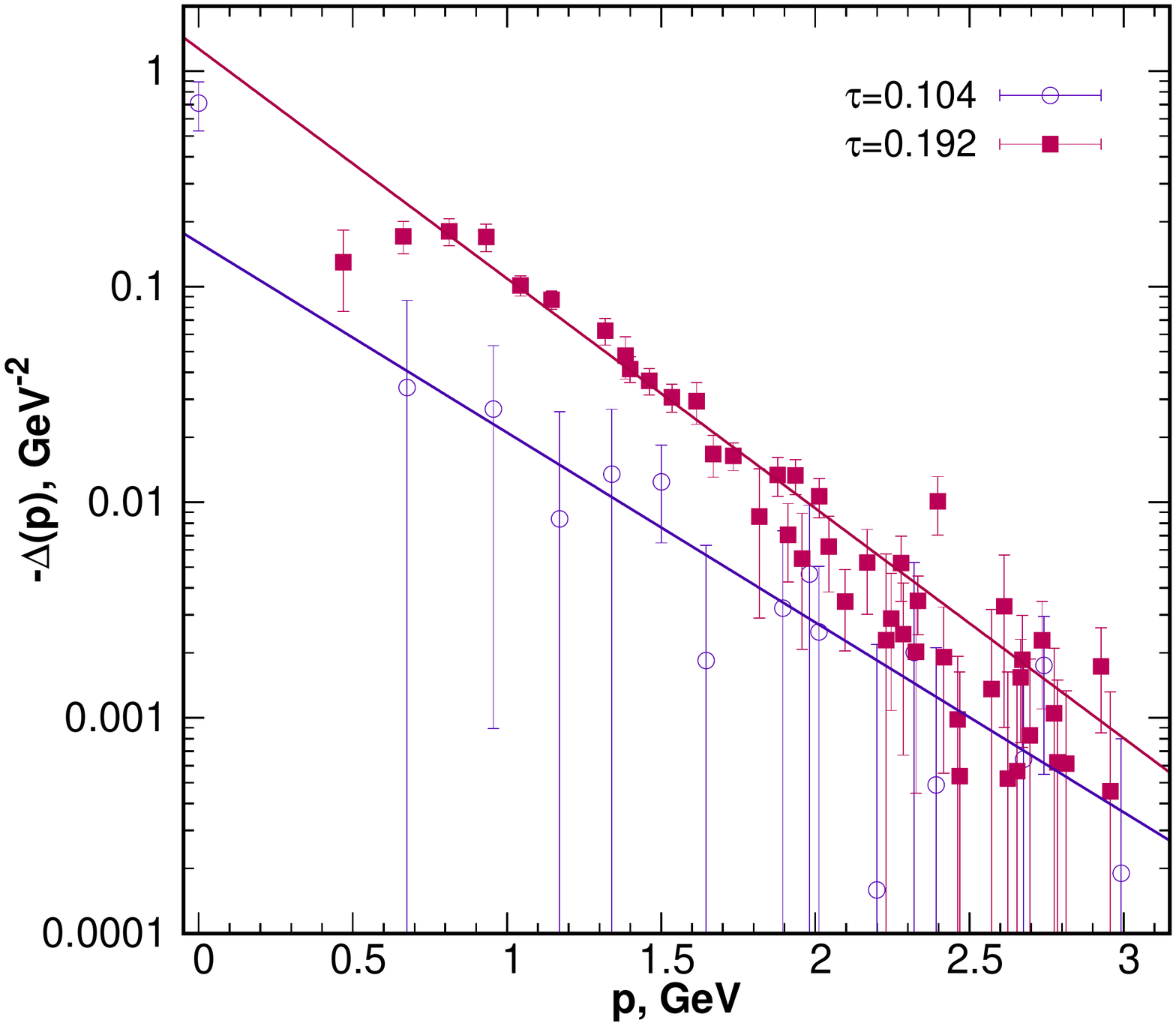}
\caption{Dependence of $\Delta$ on $|\vec p|$ in the SU(3) theory: $T<T_p$ (left panel), $T>T_p$ (right panel). Lines are the results of the fit by formula (\ref{eq:diff_basic_fit}). }
\label{fig:dglp_mitr}
\end{figure*}

Recently it was found \cite{Bornyakov:2020kyz,Bornyakov:2021mfj} that the momentum dependence of the difference between the longitudinal and transverse propagators 
$\Delta(p)=D_L(p)-D_T(p)$ in dense quark matter can well be fitted by the function
\beq\label{eq:diff_basic_fit}
\Delta(\vec p) \simeq c\exp\Big(-\nu |\vec p|\Big)
\eeq
over a sufficiently wide range of momenta.
Here we study this fit in more details and 
find temperature dependence of the fit parameters
$c$ and $\nu$ 
in SU(2) and SU(3) gluodynamics.

 In Refs.~\cite{Bornyakov:2020kyz,Bornyakov:2021mfj} 
it was shown that the Gribov-Stingl fit function
\beq\label{eq:GS_ff_DOS18}
D(p)= Z \;{M_1^2 +  p^2 \over p^4+ M_2^2 \, p^2 + M_3^4 }\; .
\eeq
works well for the longitudinal propagator, however,
a poor quality of this fit in the case of the transverse propagator was found.
Additionally it was found that the Gribov-Stingl 
fit for the transverse propagator is unstable with respect to an exclusion of the zero momentum. Thus, expression (\ref{eq:diff_basic_fit}) should be helpful for  
finding an adequate fit function for the transverse gluon propagator.

It is reasonable to determine the parameters 
$M_i^2$ and $Z$  from the fit  (\ref{eq:GS_ff_DOS18}) to $D_L(p)$,
the parameters $c$ and $\nu$ from the fit (\ref{eq:diff_basic_fit})  to $\Delta(p)$
and consider the sum of  the functions 
(\ref{eq:GS_ff_DOS18}) and (\ref{eq:diff_basic_fit})
with these parameters as an approximation to $D_T(p)$.

Typical dependence of  $\Delta$ on $|\vec p|$ in the SU(2) theory is shown in Fig.~\ref{fig:FVE}
both at $T<T_p$ (left panel) and at $T>T_p$
(right panel).
The results are presented for different lattice sizes to demonstrate that finite-volume effects in the domain of fitting are negligibly small.
In the left panel it is seen that  
the dependence of $\Delta$ on the momentum abruptly changes at $p=p_c\sim 1.2$~GeV. However, 
$\Delta(\vec p)$ can be fitted by 
eq.~(\ref {eq:diff_basic_fit}) both at $p<p_c$
and at $p>p_c$ with different sets of parameters 
$c$ and $\nu$ for different fitting ranges.

\begin{figure}[tbh]
\vspace*{-15mm}
\hspace*{-5mm}\includegraphics[width=9cm]{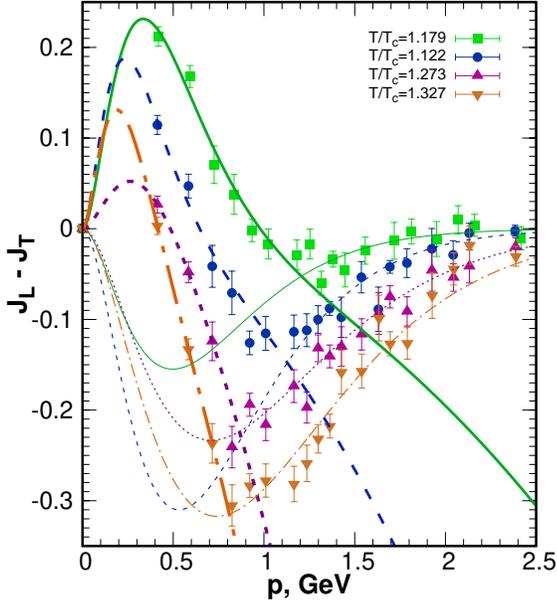}
\vspace*{-23mm}
\caption{The behavior of the difference between the longitudinal and transverse gluon dressing functions in the vicinity of $T_p$ in the SU(2) theory. Bold lines are the results of the fit by formula (\ref{eq:diff_onchange_fit})
at $p<p_c$, thin lines --- by formula (\ref{eq:diff_basic_fit}) at $p>p_c$.}
\label{fig:on_change}
\end{figure}


\begin{table}[tbh]
\begin{center}
\vspace*{0.2cm}
\begin{tabular}{|c|c|c|c|} \hline
             &         &          &                        \\[-2mm]
 ~~~$\tau$~~~&~$ln |c|$~&~~$\nu$~~&~$p$-value~ \\
             &         &          &                     \\[-2mm]
             &         &          &                        \\
   \hline\hline
-0.099 & 5.220(15) & 4.486(21) & 0.03  \\
-0.008 & 5.488(18) & 4.640(25) & 0.89  \\
-0.001 & 5.573(19) & 4.730(27) & 0.64  \\
0.002  & 5.626(20) & 4.780(25) & 0.33  \\
0.005  & 5.585(27) & 4.783(36) & 0.68  \\
0.008  & 5.727(21) & 4.850(27) & 0.98  \\
0.014  & 5.693(24) & 4.892(32) & 1.00  \\
0.025  & 5.818(29) & 5.058(40) & 1.00 \\
0.034  & 5.741(33) & 4.982(44) & 1.00 \\
0.054  & 6.177(51) & 5.318(71) & 0.71  \\
0.106  & 3.052(65) & 4.440(81) & 0.66   \\
\hline\hline
\end{tabular}
\end{center}
\caption{Parameters $c$ and $\nu$ from the fit 
formula (\ref{eq:diff_basic_fit}) for $T<T_p$
(SU(2) theory).}
\label{tab:SU2_exp_fit_p}
\end{table}

\begin{table}[tbh]
\begin{center}
\vspace*{0.2cm}
\begin{tabular}{|c|c|c|c|} \hline
                     &          &       &            \\[-2mm]
 ~~~$T/T_c$~~~  &~$ln |c|$~&~$\nu$~&~$p$-value~ \\
                     &          &       &            \\[-2mm]
                     &          &       &            \\
   \hline\hline
1.217 & 0.33(23) & 2.45(20) & 0.68  \\
1.273 & 1.05(15) & 2.63(13) & 0.38   \\
1.327 & 1.35(13) & 2.60(11) & 0.24    \\
1.494 & 1.94(5)  & 2.68(4)  & 0.82     \\
1.826 & 1.78(4)  & 2.23(4)  & 0.16   \\
2.324 & 1.58(5)  & 1.87(3)  & 0.21    \\
\hline\hline
\end{tabular}
\end{center}
\caption{Parameters $c$ and $\nu$ from the fit 
formula (\ref{eq:diff_basic_fit}) for $T>T_p$
(SU(2) theory).
}
\label{tab:SU2_exp_fit_m}
\end{table}

\begin{table}[tbh]
\begin{center}
\vspace*{0.2cm}
\begin{tabular}{|c|c|c|c|} \hline
                     &          &       &            \\[-2mm]
 ~~~$\tau$~~~  &~$c$~(GeV$^-2$)~&~$\nu$~(GeV$^-1$)~&~$p$-value~ \\
                     &          &       &            \\[-2mm]
                     &          &       &            \\
   \hline\hline
-0.096 & 61.9(5.0)  & 4.33(9)   & 0.02  \\
-0.026 & 48.7(8.6)  & 4.18(17)  & 0.01   \\
0.025  & 17.3(2.7)  & 4.15(13)  & 0.80    \\
0.104  & -16.5(2.2) & 4.10(12)  & 0.79     \\
\hline\hline
\end{tabular}
\end{center}
\caption{Parameters $c$ and $\nu$ from the fit 
formula (\ref{eq:diff_basic_fit}) in the SU(3) theory.
}
\label{tab:SU3_exp_fit}
\end{table}

Similar results for the SU(3) theory are presented in Fig~\ref{fig:dglp_mitr}.

In the case of SU(2) theory, the behavior of $\Delta(\vec p)$ at $T\sim T_p$
is shown in Fig.~\ref{fig:on_change}.
To make it visible in the plot,
we use the difference $J_L-J_T$ between
the dressing functions $J_{L,T}(p)= p^2 D_{L,T}(p)$
instead of $\Delta(\vec p)$.

To describe the behavior of $\Delta(\vec p)$
at $T\sim T_p$, we employ fit formula 
\beq\label{eq:diff_onchange_fit}
\Delta(\vec p) \simeq c\exp\Big(-\nu |\vec p|\Big) - b
\eeq
at $p<p_c$ and (\ref{eq:diff_basic_fit}) at $p>p_c$.

The parameters $c$ and $\nu$ obtained by
fitting the formula (\ref{eq:diff_basic_fit})
to the data are given in
Tables~\ref{tab:SU2_exp_fit_p}
and~\ref{tab:SU2_exp_fit_m} for SU(2) theory and 
in Table~\ref{tab:SU3_exp_fit} for SU(3) theory.

In the case of SU(2) the results for $T> T_p$ and $T< T_p$ are shown separately because at $T<T_p$
we use the range of fitting $p<p_c$,
whereas $p_c$ decreases at $T>T_c$  and we 
consider the range of fitting at $p>p_c$
as the main relevant.
The range of fitting employed for these tables 
is $0.3$~GeV$<|\vec p|<1.1$~GeV 
at $T<T_p$ and $1.0$~GeV$<|\vec p|<3\div 4$~GeV 
at $T>T_p$. The results at the temperatures close of $T_p$ are not presented in these tables
because the behavior of $\Delta(\vec p)$
changes and the domain of validity of the fit formula (\ref{eq:diff_basic_fit}) associated with an appropriate range of fitting also changes
as is discussed in connection with Fig~\ref{fig:on_change}.

\begin{figure*}[tbh]
\vspace*{-16mm}
\hspace*{-5mm}\includegraphics[width=9cm]{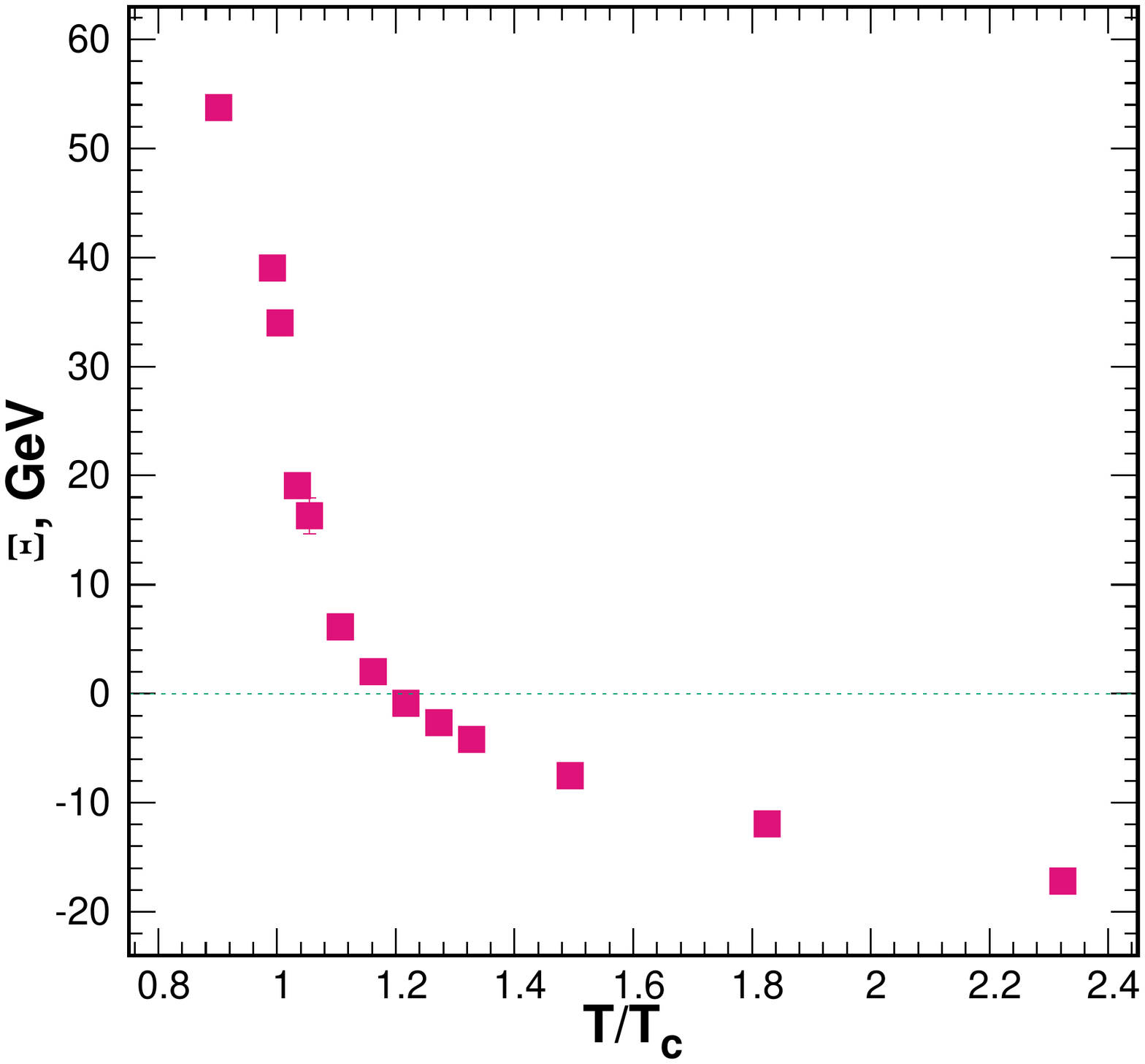}\hspace*{-8mm}
\includegraphics[width=9cm]{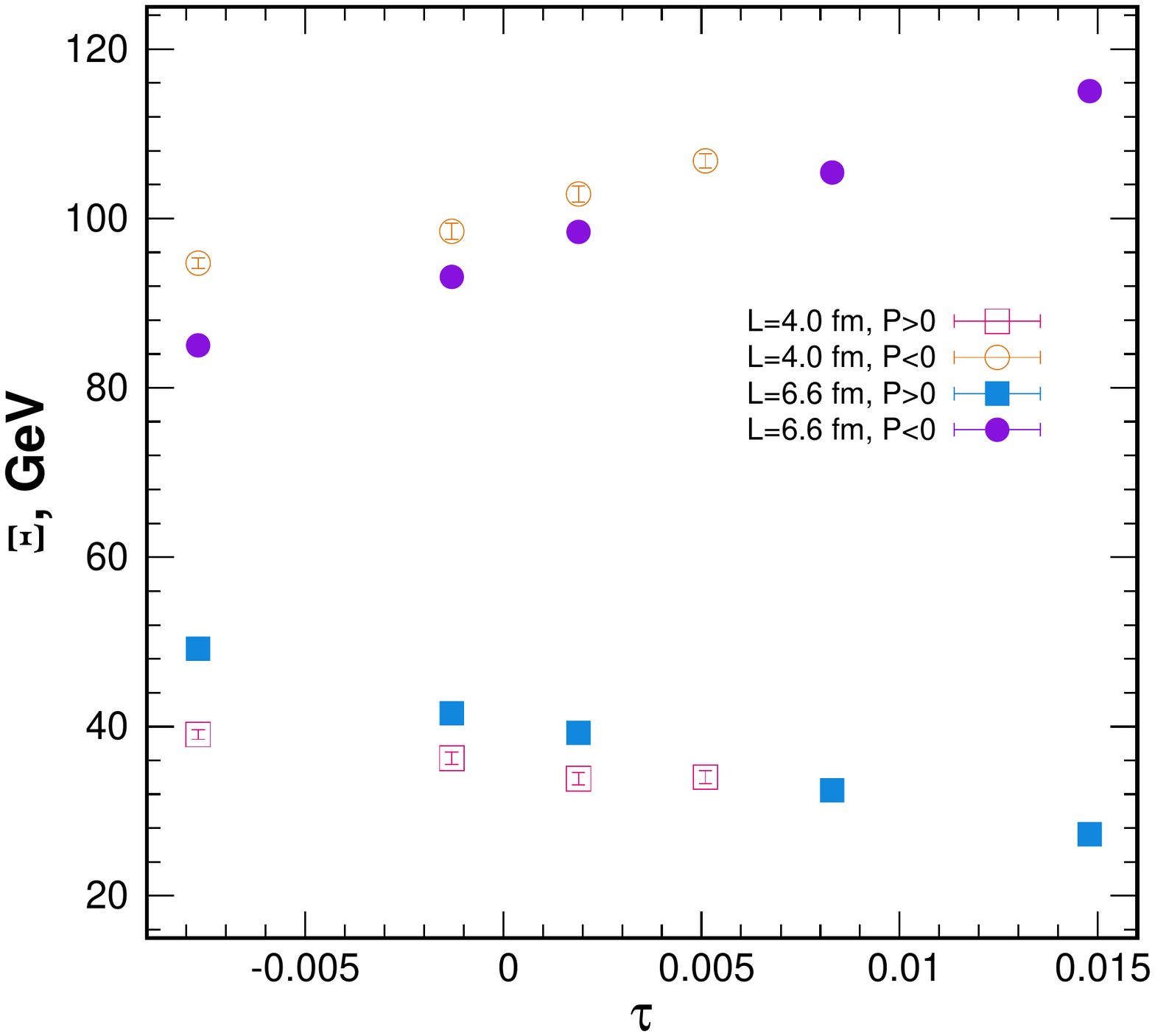}
\vspace*{-25mm}
\caption{Left panel: temperature dependence of $\Xi$ in the SU(2) theory over the entire temperature range under consideration (for the sector ${\cal P}>0$). Right panel: temperature dependence of $\Xi$ at $T\approx T_c$ of $\Xi$ in different Polyakov-loop sectors and on a lattices of different size. }
\label{fig:Xi_vs_T_SU2}
\end{figure*}

\begin{figure}[tbh]
\vspace*{1mm}
\hspace*{-5mm}\includegraphics[width=9cm]{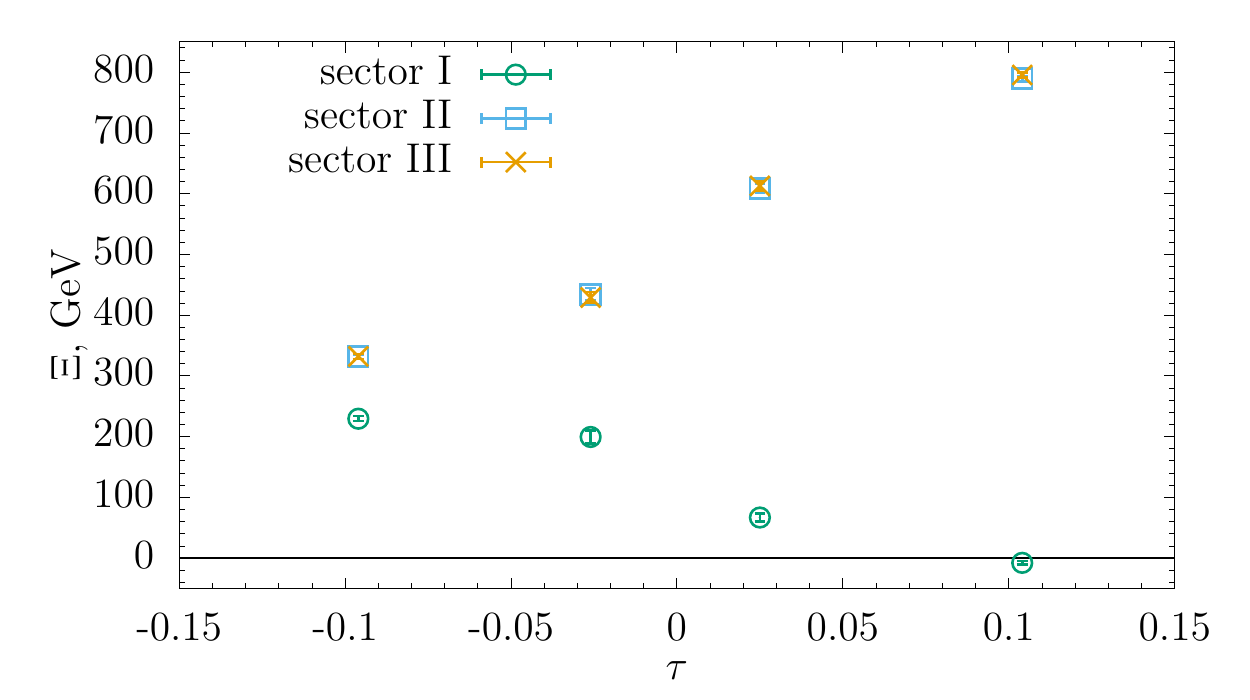}
\caption{Temperature dependence of $\Xi$ in different center sectors for SU(3) theory at $T\sim T_c$.}
\label{fig:SU3-Xi_tau}
\end{figure}

\begin{figure}[h]
\vspace*{1mm}
\includegraphics[width=8cm]{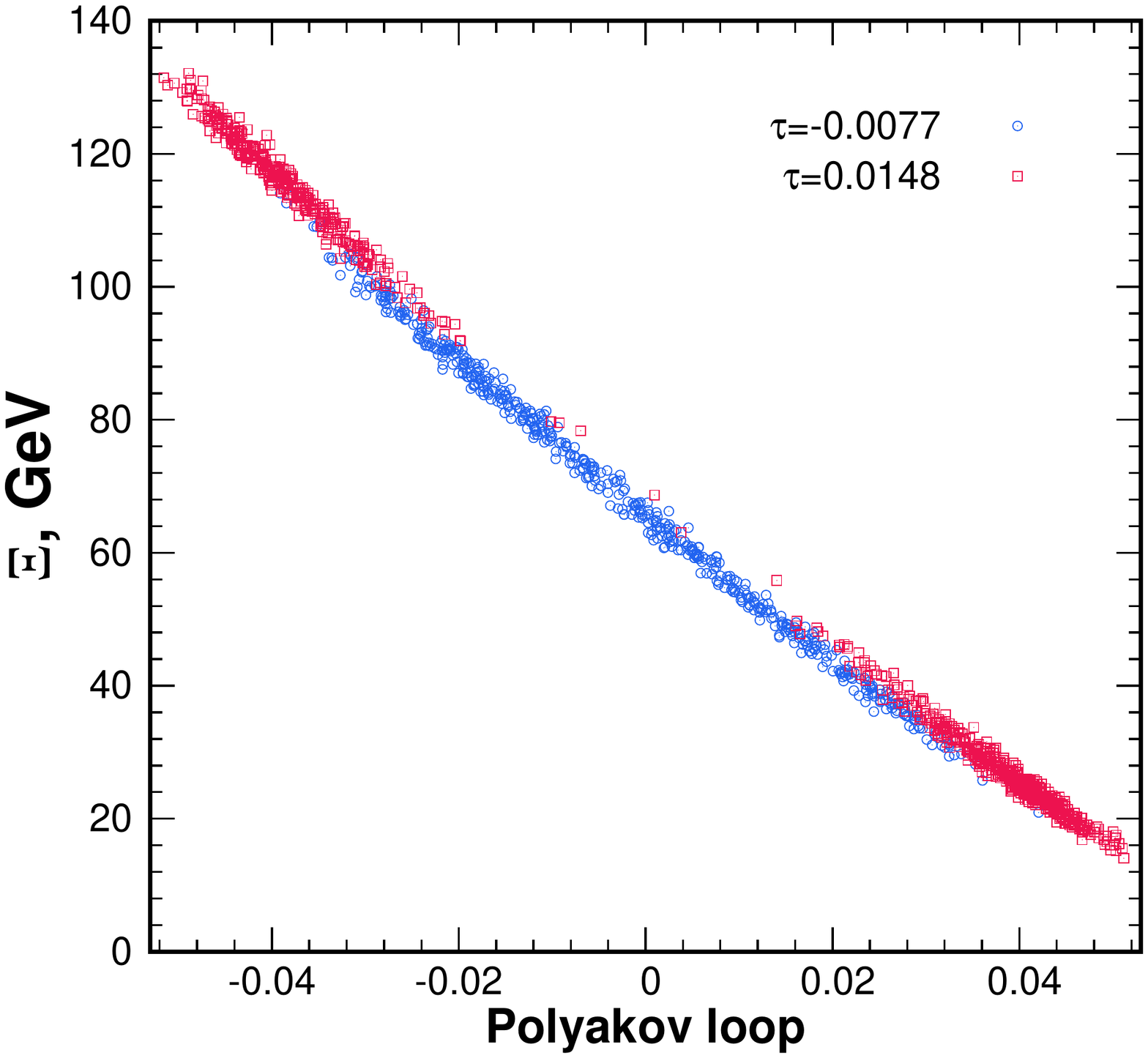}\\
\vspace*{-17ex}\includegraphics[width=8cm]{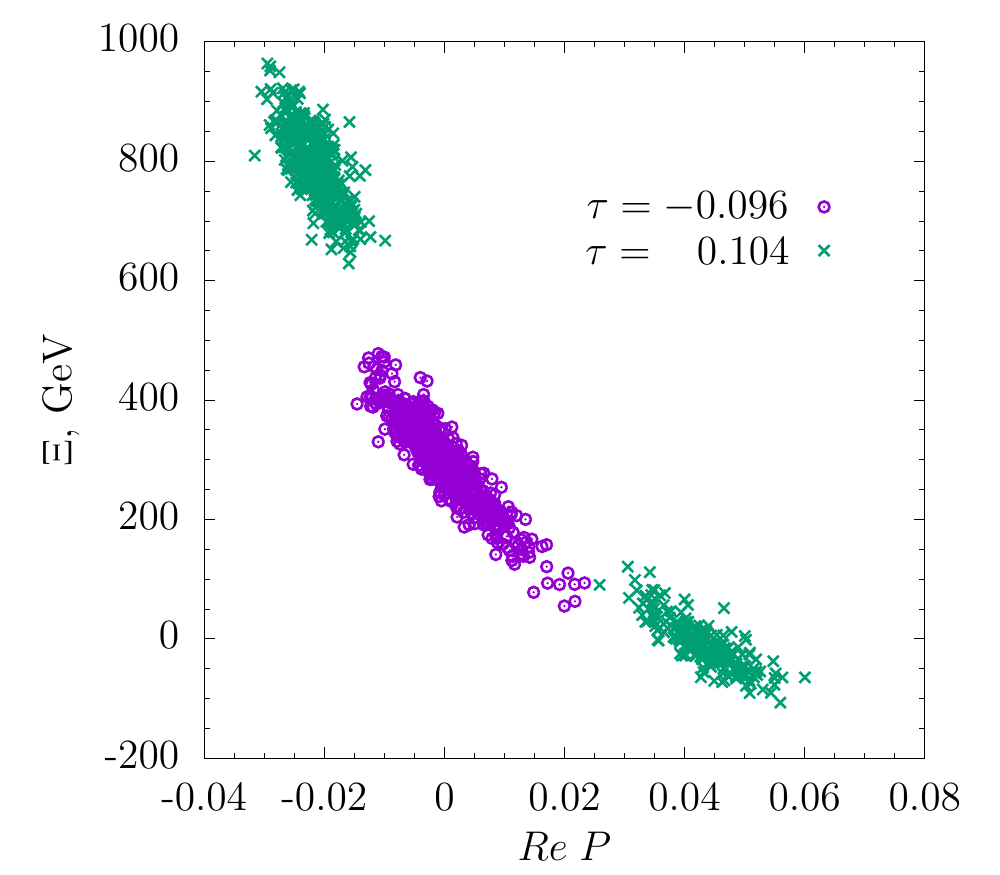}
\caption{Correlations between $\Xi$ and the Polyakov loop: SU(2) theory (upper panel) and SU(3) theory (lower panel).}
\label{fig:scatter_plots}
\end{figure}

\section{Integrated difference of the propagators}


Since the difference between the propagators
$\Delta(p)$ decreases rapidly with the momentum,
it is reasonable to perform its integration 
with respect to the momentum 
in order to obtain a quantity 
sensitive to infrared dynamics
of gauge fields. It should be emphasized that
the asymmetry $\aasymt$ can hardly be an indicator of infrared dynamics 
because it receives contributions from all momenta. 
Moreover, at $p_4=0$ the contribution of high momenta dominates over 
the contribution of low momenta. The ultraviolet convergence of $\aasymt$ stems from 
cancellation of the high-$|\vec p|$ 
contributions at low $p_4$ and 
the contributions of high $p_4$. 

 
In situations when the concept 
of potential is relevant
and the interaction potential 
can be characterized by 
the Fourier transform of the propagator,
the depth of the potential well is associated with
the integral of the propagator over all momenta.
Therefore, the integral of $\Delta(\vec p,0)$
with respect to the 3-momentum 
characterizes the difference between the 
potentials 
of chromoelectric and chromomagnetic interactions.

Thus we define the Integrated Difference between the longitudinal and transverse Propagators (IDP)
as follows
\beq\label{eq:IDP_def}
\Xi=\int d\vec p  \big( D_L(\vec p, 0)-D_T(\vec p, 0) \big) \;.
\eeq
It describes an overall contribution 
of infrared gluon dynamics 
to the difference between chromoelectric and chromomagnetic interactions.
The IDP $\Xi$ is readily rearranged to the form
\bea
&&\Xi = {(2\pi)^3 T \over V (N_c^2-1)} \int d\vec x 
\int_0^{1/T}\hspace*{-3mm} d x_4  \int_0^{1/T} \hspace*{-3mm} d y_4 \\ \nonumber
&& \sum_{b=0}^{N_c} \Big\langle A_4^b(\vec x, x_4) A_4^b(\vec x, y_4) -
{1\over 2} \sum_{i=1}^3 A_i^b(\vec x, x_4) A_i^b(\vec x, y_4) \Big\rangle \;, \nonumber
\eea 
it can also be expressed in terms of lattice variables as follows: 
\bea
 &&   \Xi = {16 \pi^3 \beta T \over N_s^3 N_c(N_c^2-1)}\; \sum_{\vec x}\! \sum_{x_4, y_4}\! \sum_{b=1}^3 \\  \nonumber 
&& \left( \big\langle u_4^b(\vec x,x_4) u_4^b(\vec x,y_4)\big\rangle - {1\over 2} \sum_{i=1}^3 \; \big\langle u_i^b(\vec x,x_4) u_i^b(\vec x,y_4)\big\rangle \right) \nonumber 
\eea
$x=(\vec x,x_4)\in \Lambda$ --- $N_s^3\times N_t$ lattice.

In the free theory $\Delta=0$ and, therefore,
 $\Xi=0$. This contrasts with the asymmetry
 $\aasymt$ (\ref{eq:asym_thr_prop}), which in the free theory on a lattice can be recast to the form
\beq\label{eq:asym_free}
\aasymt={2N_c (N_c^2-1) N_t\over 3 \beta a^2 N_s^3} \sum_{\hat p} { |\hat{\vec p}|^2 - 3 \hat p_4^2 \over 3 (\hat p^2)^2}
\eeq
As is seen from this formula, 
the contribution of the $p_4=0$ mode 
to the asymmetry diverges when $a\to 0$ 
at high $\hat p_i$ and only the contributions 
of high-$p_4$ modes makes the asymmetry finite
$\ds \aasymt\sim \;-\;{1\over 12}$ when $a\to 0$ \cite{Chernodub:2008kf}.
For this reason, the IDP is better suited 
to characterize strength of interactions 
than the asymmetry, 
especially in the infrared domain.
Yet another important difference between the 
IDP and the asymmetry is that $\Xi(T)$
goes through zero at some temperature,
whereas $\aasymt$ does not \cite{Bornyakov:2016geh}.

Temperature dependence of $\Xi$ in SU(2) theory
in the center sector characterized by 
the positive values of the Polyakov loop
is shown on the left panel of Fig.~\ref{fig:Xi_vs_T_SU2}.
It is clear that it goes through zero 
at $T\approx 1.2 T_c$. 
At this temperature the longitudinal propagator associated with chromoelectric interactions 
becomes smaller than the transverse propagator  associated with chromomagnetic interactions.
This is not true for zero-momentum values of these
propagators: $D_L(0)>D_T(0)$ even 
at substantially greater temperatures, however, 
the leading contribution to $\Xi$ comes from momenta 
$|\vec p|> 200\div 400$~MeV, where $D_T>D_L$ at $T> T_p\approx 1.2T_c$. It should also be mentioned that 
$D_T(0)$ is plagued by the finite-volume and 
Gribov-copy effects and this fact hinders the
determination of the temperature where $D_L(0)=D_T(0)$.
Fortunately, the determination of such point is 
not very important since we have a useful quantity
$\Xi$ reflecting the interplay between chromoelectric and chromomagnetic forces.
At $T< T_p$ chromoelectric forces dominate: 
they are far-distant and sufficiently strong,
which characterizes the postconfinement domain.
However, with an increase of the chromoelectric mass
they become screened at short distances
so that chromomagnetic forces become dominating
and we arrive at deconfined  gluon matter.
Thus $T_p$ should be considered as a natural boundary of the postconfinement domain.
However, the question remains about gauge dependence of $T_p$. 

It should be emphasized that $\Xi$ depends 
significantly on the center sector, 
its temperature dependence in different
center sectors is shown in the right panel of 
Fig.~\ref{fig:Xi_vs_T_SU2}. 

A similar pattern takes place in the case of SU(3),
it is shown in Fig.~\ref{fig:SU3-Xi_tau}.
In this case we obtain $T_p\approx 1.1 T_c$.
This value of the boundary of the postconfinement domain is substantially lower than that in \cite{Dumitru:2010mj,Hidaka:2015ima}.

We also observe a significant correlation between $\Xi$ and ${\cal P}$, which is shown by the scatter plot in Fig.~\ref{fig:scatter_plots}. 
Recently it was argued \cite{Bornyakov:2021pls} that
such correlations imply similarity of critical behavior of the correlated quantities in the infinite-volume limit.  
Therefore, we expect that $\Xi(T)\sim \tau^\beta$ as $\tau\to 0_+$ in the case of SU(2) and $\Xi(T)$ has a discontinuity 
at $T=T_c$.

\section{Conclusions}

We have studied the momentum and temperature 
dependence of the difference between the longitudinal and transverse propagators $\Delta(\vec p, p_4=0;T)$.

Our findings can be summarized as follows:

\begin{itemize}
\item In a sufficiently wide range 
of infrared momenta $\Delta(\vec p)$ 
can well be fitted 
by the function (\ref{eq:diff_basic_fit}) 
and the parameter $c$ changes its sign 
at $T\simeq 1.2T_c$ in the SU(2) theory 
and at $T\simeq 1.1T_c$ in the SU(3) theory.
\item In the center sector with 
a positive real part of the Polyakov-loop, 
the integrated difference of propagators $\Xi$
goes through zero at $T\simeq 1.2T_c$ 
in the SU(2) theory 
and at $T\simeq 1.1T_c$ in the SU(3) theory.
\item The temperature $T_p$ at which $\Xi$ goes 
through zero
can be considered as the boundary of the postconfinement domain, where chromoelectric interactions still dominate.
\item A significant correlation between $\Xi$ and $\mathrm{Re}{\cal P}$ is observed indicating that these quantities have similar critical behavior.
\end{itemize}

\vspace*{2mm}

\acknowledgments{Computer simulations were performed on the IHEP 
Central Linux Cluster and ITEP Linux Cluster.
This work was supported by the Russian Foundation for Basic Research, 
grant~no.20-02-00737~A.}

\bibliographystyle{apsrev}
\bibliography{citations_asym_2016}

\begin{thebibliography}{17}
\expandafter\ifx\csname natexlab\endcsname\relax\def\natexlab#1{#1}\fi
\expandafter\ifx\csname bibnamefont\endcsname\relax
  \def\bibnamefont#1{#1}\fi
\expandafter\ifx\csname bibfnamefont\endcsname\relax
  \def\bibfnamefont#1{#1}\fi
\expandafter\ifx\csname citenamefont\endcsname\relax
  \def\citenamefont#1{#1}\fi
\expandafter\ifx\csname url\endcsname\relax
  \def\url#1{\texttt{#1}}\fi
\expandafter\ifx\csname urlprefix\endcsname\relax\def\urlprefix{URL }\fi
\providecommand{\bibinfo}[2]{#2}
\providecommand{\eprint}[2][]{\url{#2}}

\bibitem[{\citenamefont{Bazavov et~al.}(2012)}]{Bazavov:2011nk}
\bibinfo{author}{\bibfnamefont{A.}~\bibnamefont{Bazavov}} \bibnamefont{et~al.},
  \bibinfo{journal}{Phys. Rev. D} \textbf{\bibinfo{volume}{85}},
  \bibinfo{pages}{054503} (\bibinfo{year}{2012}), \eprint{1111.1710}.

\bibitem[{\citenamefont{Bazavov et~al.}(2014)}]{HotQCD:2014kol}
\bibinfo{author}{\bibfnamefont{A.}~\bibnamefont{Bazavov}} \bibnamefont{et~al.}
  (\bibinfo{collaboration}{HotQCD}), \bibinfo{journal}{Phys. Rev. D}
  \textbf{\bibinfo{volume}{90}}, \bibinfo{pages}{094503}
  (\bibinfo{year}{2014}), \eprint{1407.6387}.

\bibitem[{\citenamefont{Kaczmarek et~al.}(2002)\citenamefont{Kaczmarek, Karsch,
  Petreczky, and Zantow}}]{Kaczmarek:2002mc}
\bibinfo{author}{\bibfnamefont{O.}~\bibnamefont{Kaczmarek}},
  \bibinfo{author}{\bibfnamefont{F.}~\bibnamefont{Karsch}},
  \bibinfo{author}{\bibfnamefont{P.}~\bibnamefont{Petreczky}},
  \bibnamefont{and} \bibinfo{author}{\bibfnamefont{F.}~\bibnamefont{Zantow}},
  \bibinfo{journal}{Phys. Lett.} \textbf{\bibinfo{volume}{B543}},
  \bibinfo{pages}{41} (\bibinfo{year}{2002}), \eprint{hep-lat/0207002}.

\bibitem[{\citenamefont{Dumitru et~al.}(2004)\citenamefont{Dumitru, Hatta,
  Lenaghan, Orginos, and Pisarski}}]{Dumitru:2003hp}
\bibinfo{author}{\bibfnamefont{A.}~\bibnamefont{Dumitru}},
  \bibinfo{author}{\bibfnamefont{Y.}~\bibnamefont{Hatta}},
  \bibinfo{author}{\bibfnamefont{J.}~\bibnamefont{Lenaghan}},
  \bibinfo{author}{\bibfnamefont{K.}~\bibnamefont{Orginos}}, \bibnamefont{and}
  \bibinfo{author}{\bibfnamefont{R.~D.} \bibnamefont{Pisarski}},
  \bibinfo{journal}{Phys. Rev.} \textbf{\bibinfo{volume}{D70}},
  \bibinfo{pages}{034511} (\bibinfo{year}{2004}), \eprint{hep-th/0311223}.

\bibitem[{\citenamefont{Dumitru et~al.}(2011)\citenamefont{Dumitru, Guo,
  Hidaka, Altes, and Pisarski}}]{Dumitru:2010mj}
\bibinfo{author}{\bibfnamefont{A.}~\bibnamefont{Dumitru}},
  \bibinfo{author}{\bibfnamefont{Y.}~\bibnamefont{Guo}},
  \bibinfo{author}{\bibfnamefont{Y.}~\bibnamefont{Hidaka}},
  \bibinfo{author}{\bibfnamefont{C.~P.~K.} \bibnamefont{Altes}},
  \bibnamefont{and} \bibinfo{author}{\bibfnamefont{R.~D.}
  \bibnamefont{Pisarski}}, \bibinfo{journal}{Phys. Rev.}
  \textbf{\bibinfo{volume}{D83}}, \bibinfo{pages}{034022}
  (\bibinfo{year}{2011}), \eprint{1011.3820}.

\bibitem[{\citenamefont{Hidaka et~al.}(2015)\citenamefont{Hidaka, Lin,
  Pisarski, and Satow}}]{Hidaka:2015ima}
\bibinfo{author}{\bibfnamefont{Y.}~\bibnamefont{Hidaka}},
  \bibinfo{author}{\bibfnamefont{S.}~\bibnamefont{Lin}},
  \bibinfo{author}{\bibfnamefont{R.~D.} \bibnamefont{Pisarski}},
  \bibnamefont{and} \bibinfo{author}{\bibfnamefont{D.}~\bibnamefont{Satow}},
  \bibinfo{journal}{JHEP} \textbf{\bibinfo{volume}{10}}, \bibinfo{pages}{005}
  (\bibinfo{year}{2015}), \eprint{1504.01770}.

\bibitem[{\citenamefont{Asakawa and Hatsuda}(2004)}]{Asakawa:2003re}
\bibinfo{author}{\bibfnamefont{M.}~\bibnamefont{Asakawa}} \bibnamefont{and}
  \bibinfo{author}{\bibfnamefont{T.}~\bibnamefont{Hatsuda}},
  \bibinfo{journal}{Phys. Rev. Lett.} \textbf{\bibinfo{volume}{92}},
  \bibinfo{pages}{012001} (\bibinfo{year}{2004}), \eprint{hep-lat/0308034}.

\bibitem[{\citenamefont{Chernodub and Ilgenfritz}(2008)}]{Chernodub:2008kf}
\bibinfo{author}{\bibfnamefont{M.~N.} \bibnamefont{Chernodub}}
  \bibnamefont{and} \bibinfo{author}{\bibfnamefont{E.~M.}
  \bibnamefont{Ilgenfritz}}, \bibinfo{journal}{Phys. Rev.}
  \textbf{\bibinfo{volume}{D78}}, \bibinfo{pages}{034036}
  (\bibinfo{year}{2008}), \eprint{0805.3714}.

\bibitem[{\citenamefont{Bornyakov et~al.}(2016)\citenamefont{Bornyakov,
  Mitrjushkin, and Rogalyov}}]{Bornyakov:2016geh}
\bibinfo{author}{\bibfnamefont{V.~G.} \bibnamefont{Bornyakov}},
  \bibinfo{author}{\bibfnamefont{V.~K.} \bibnamefont{Mitrjushkin}},
  \bibnamefont{and} \bibinfo{author}{\bibfnamefont{R.~N.}
  \bibnamefont{Rogalyov}} (\bibinfo{year}{2016}), \eprint{1609.05145}.

\bibitem[{\citenamefont{Bornyakov and Mitrjushkin}(2011)}]{Bornyakov:2011jm}
\bibinfo{author}{\bibfnamefont{V.~G.} \bibnamefont{Bornyakov}}
  \bibnamefont{and} \bibinfo{author}{\bibfnamefont{V.~K.}
  \bibnamefont{Mitrjushkin}} (\bibinfo{year}{2011}), \eprint{1103.0442}.

\bibitem[{\citenamefont{Aouane et~al.}(2012)\citenamefont{Aouane, Bornyakov,
  Ilgenfritz, Mitrjushkin, Muller-Preussker et~al.}}]{Aouane:2011fv}
\bibinfo{author}{\bibfnamefont{R.}~\bibnamefont{Aouane}},
  \bibinfo{author}{\bibfnamefont{V.}~\bibnamefont{Bornyakov}},
  \bibinfo{author}{\bibfnamefont{E.}~\bibnamefont{Ilgenfritz}},
  \bibinfo{author}{\bibfnamefont{V.}~\bibnamefont{Mitrjushkin}},
  \bibinfo{author}{\bibfnamefont{M.}~\bibnamefont{Muller-Preussker}},
  \bibnamefont{et~al.}, \bibinfo{journal}{Phys.Rev.}
  \textbf{\bibinfo{volume}{D85}}, \bibinfo{pages}{034501}
  (\bibinfo{year}{2012}), \eprint{1108.1735}.

\bibitem[{\citenamefont{Necco and Sommer}(2002)}]{Necco:2001xg}
\bibinfo{author}{\bibfnamefont{S.}~\bibnamefont{Necco}} \bibnamefont{and}
  \bibinfo{author}{\bibfnamefont{R.}~\bibnamefont{Sommer}},
  \bibinfo{journal}{Nucl. Phys. B} \textbf{\bibinfo{volume}{622}},
  \bibinfo{pages}{328} (\bibinfo{year}{2002}), \eprint{hep-lat/0108008}.

\bibitem[{\citenamefont{Boyd et~al.}(1996)\citenamefont{Boyd, Engels, Karsch,
  Laermann, Legeland, Lutgemeier, and Petersson}}]{Boyd:1996bx}
\bibinfo{author}{\bibfnamefont{G.}~\bibnamefont{Boyd}},
  \bibinfo{author}{\bibfnamefont{J.}~\bibnamefont{Engels}},
  \bibinfo{author}{\bibfnamefont{F.}~\bibnamefont{Karsch}},
  \bibinfo{author}{\bibfnamefont{E.}~\bibnamefont{Laermann}},
  \bibinfo{author}{\bibfnamefont{C.}~\bibnamefont{Legeland}},
  \bibinfo{author}{\bibfnamefont{M.}~\bibnamefont{Lutgemeier}},
  \bibnamefont{and}
  \bibinfo{author}{\bibfnamefont{B.}~\bibnamefont{Petersson}},
  \bibinfo{journal}{Nucl. Phys. B} \textbf{\bibinfo{volume}{469}},
  \bibinfo{pages}{419} (\bibinfo{year}{1996}), \eprint{hep-lat/9602007}.

\bibitem[{\citenamefont{Fingberg et~al.}(1993)\citenamefont{Fingberg, Heller,
  and Karsch}}]{Fingberg:1992ju}
\bibinfo{author}{\bibfnamefont{J.}~\bibnamefont{Fingberg}},
  \bibinfo{author}{\bibfnamefont{U.~M.} \bibnamefont{Heller}},
  \bibnamefont{and} \bibinfo{author}{\bibfnamefont{F.}~\bibnamefont{Karsch}},
  \bibinfo{journal}{Nucl. Phys.} \textbf{\bibinfo{volume}{B392}},
  \bibinfo{pages}{493} (\bibinfo{year}{1993}), \eprint{hep-lat/9208012}.

\bibitem[{\citenamefont{Bornyakov et~al.}(2020)\citenamefont{Bornyakov,
  Braguta, Nikolaev, and Rogalyov}}]{Bornyakov:2020kyz}
\bibinfo{author}{\bibfnamefont{V.~G.} \bibnamefont{Bornyakov}},
  \bibinfo{author}{\bibfnamefont{V.~V.} \bibnamefont{Braguta}},
  \bibinfo{author}{\bibfnamefont{A.~A.} \bibnamefont{Nikolaev}},
  \bibnamefont{and} \bibinfo{author}{\bibfnamefont{R.~N.}
  \bibnamefont{Rogalyov}}, \bibinfo{journal}{Phys. Rev. D}
  \textbf{\bibinfo{volume}{102}}, \bibinfo{pages}{114511}
  (\bibinfo{year}{2020}), \eprint{2003.00232}.

\bibitem[{\citenamefont{Bornyakov
  et~al.}(2021{\natexlab{a}})\citenamefont{Bornyakov, Nikolaev, Rogalyov, and
  Terentev}}]{Bornyakov:2021mfj}
\bibinfo{author}{\bibfnamefont{V.~G.} \bibnamefont{Bornyakov}},
  \bibinfo{author}{\bibfnamefont{A.~A.} \bibnamefont{Nikolaev}},
  \bibinfo{author}{\bibfnamefont{R.~N.} \bibnamefont{Rogalyov}},
  \bibnamefont{and} \bibinfo{author}{\bibfnamefont{A.~S.}
  \bibnamefont{Terentev}}, \bibinfo{journal}{Eur. Phys. J. C}
  \textbf{\bibinfo{volume}{81}}, \bibinfo{pages}{747}
  (\bibinfo{year}{2021}{\natexlab{a}}), \eprint{2102.07821}.

\bibitem[{\citenamefont{Bornyakov
  et~al.}(2021{\natexlab{b}})\citenamefont{Bornyakov, Goy, Mitrjushkin, and
  Rogalyov}}]{Bornyakov:2021pls}
\bibinfo{author}{\bibfnamefont{V.~G.} \bibnamefont{Bornyakov}},
  \bibinfo{author}{\bibfnamefont{V.~A.} \bibnamefont{Goy}},
  \bibinfo{author}{\bibfnamefont{V.~K.} \bibnamefont{Mitrjushkin}},
  \bibnamefont{and} \bibinfo{author}{\bibfnamefont{R.~N.}
  \bibnamefont{Rogalyov}}, \bibinfo{journal}{Phys. Rev. D}
  \textbf{\bibinfo{volume}{104}}, \bibinfo{pages}{074508}
  (\bibinfo{year}{2021}{\natexlab{b}}), \eprint{2101.03605}.

\end{thebibliography}

\end{document}